\documentclass{PoS}
\title{Inflation, Gravitino and Reheating in Modified Modular invariant Supergravity} 
\ShortTitle{Modified Modular invariant Supergravity}
\author{\speaker{Mitsuo J. Hayashi}\\
        Department of Physics, Tokai University, Hiratsuka, 259-1292, Japan\\
        E-mail: mhayashi@keyaki.cc.u-tokai.ac.jp}

\author{Yuta Koshimizu, Toyokazu Fukuoka, Kenji Takagi, Hikoya Kasari\\
Tokai University}

\abstract{A new modified string-inspired modular invariant 
supergravity model is proposed and is applied to realize the slow roll inflation in Einstein frame.
Because inflation deals with Planck scale physics, the dilaton can be a strong candidate for identification with the inflaton.
The model we used, cleared the $\eta$-problem and negative energy problem of potential at stable point, and appeared to predict successfully the values of observations at the inflation era. 
 We proved that the model explains WMAP observations appropriately. Moreover, a mechanism of SSB and Gravitino production just after the end of inflation is investigated.
We have obtained power spectrum of the density perturbation as $\mathcal{P_R}_* \sim 2.438 \times 10^{-9}$
and the scalar spectral index as $n_{s*}\sim 0.9746$ and its tilt as $\alpha_{s*}\sim -4.3 \times 10^{-4}$. 
The ratio between scalar power spectrum and tensor is predicted as $r\sim 6.8 \times 10^{-2}$,  a prediction which seems in the range possibly observed by the Planck satellite soon. 
The gravitino mass and their production rate from scalar fields are estimated at certain values of parameters in the model.
The reheating temperature is estimated by the stability condition of Boltzmann equation by using the decay rate of the dilaton S into gauginos as $T_R({\rm gaugino}) = 3.88 \times 10^{7} \,\, {\rm GeV}$.
Though only one example of parameter choices has been  discussed here, 
the other seven candidates of parameter choices that are compatible with WMAP data  have already been found by the authors.  
Some of the examples show that the gauginos can be observed by LHC experiments.
The plausible supergravity model of inflation which here we described will open the hope to construct a realistic theory of particle theory and cosmology in this framework.}

\FullConference{35th International Conference of High Energy Physics - ICHEP2010,\\
July 22-28, 2010 \\ Paris France}

\begin{document}

N=1, d = 4 supergravity from d = 10 heterotic string by dimensional reduction has No-scale structure with $E_8 \times E_8$ gauge group \cite{01}. 
We would like to propose a new Modular invariant $N=1$ Supergravity, where the K\"{a}hler potential and thesuperpotential are given as:
$$
\!\!\!\!\!\!\!K=-\ln \!\left(S+S^\ast\right)
-3\ln \!\left(T+T^\ast-|Y|^2 \right), \qquad\quad
W=\alpha+\beta S+3bY^3\ln\left[c\>e^{S/3b}\>Y\eta^2(T)\right],
$$
where $\eta$ is  Dedekind's  $\eta$ function, $c$ is a free parameter in the theory and $S$ is a dilaton, $T$ is a moduli and $Y$ is a complex scalar superfield defined by the gaugino condensation $U\sim <\lambda\lambda>=Y^3$ of the $E_8$ hidden sector\cite{02}, and $\alpha$, $\beta$ are new parameters that should be determined from observations. 
The renormalization group parameter $b=\frac{15}{16\pi^2}$ can correspond to the $E_8$ hidden sector gauge group. \\
A modified string-inspired modular invariant supergravity is proposed here to apply it to inflationary cosmology.
Because inflation is concerned with Planck scale physics, the dilaton can be one of the strong candidates for the inflaton\cite{03, 04}.
We assume that the massless Goldstino is identified with the dilatino $\tilde{S}$ because the mass of  $\tilde{S}$  satisfies 
$m_{SS}=0$,
where $m$ is defined by $m \equiv e^{K/2} W$. \\
Then the scalar potential ($V_E \equiv e^G \left[ G_i G^{ij^*} G_{j^*} - 3 \right] $) is in order:
\begin{eqnarray}
&&\hspace{-30pt} V_E= \frac{1}{(S+S^*)(T+T^*-|Y|^2)^2}  \left[ 3 b^2 |Y|^4 \left| 1 + 3 \ln [O] \right|^2 \right. \nonumber \\[5pt]
&&\hspace{-15pt} + \frac{1}{T+T^*-|Y|^2} \left| \strut \alpha + \beta S + 3 b Y^3 \ln [O] - (S+S^*) (Y^3 + \beta) \right|^2  \nonumber \\[5pt]
&&\hspace{-15pt} + 6 b^2 |Y|^6 \left\{ \left( 1- \frac{\alpha + \beta S^*}{b{Y^*}^3} \right) \frac{ \eta' (T)}{ \eta (T)} + \left( 1- \frac{\alpha + \beta S}{bY^3} \right) \frac{ \eta' (T^*)}{ \eta (T^*)}   \right.  \left. \left. + 2 (T+T^*) \left| \frac{ \eta' (T)}{ \eta (T)} \right|^2 \right\} \right],
\end{eqnarray}
where $O=c\>e^{S/3b}\>Y\eta^2(T)$ and the potential is explicitly modular invariant in $T$. 
Instead of imposing $W_Y+K_YW=0$, we will assume $W_Y=0$ which is a rather good approximation. 
Then, a relation between $S$ and $Y$ is obtained as follows:
$Y = \frac{1}{c \eta^2 (T) e^{\frac{1}{3}}} e^{-\frac{S}{3b}}.$\\
We will here only present one case among the parameter choices $c$, $\alpha$ and $\beta$, for which the potential $V(S,Y)$ at $T=1$ has a stable minimum.  \\
Hereafter we fix 
$T=1$ ($\eta (1) = 0.768225$, $\eta^2 (1) = 0.590170$, $\eta' (1) = -0.192056$, $\eta'' (1) = -0.00925929$) and $b=\frac{15}{16\pi^2}$ corresponding to the $E_8$ gauge group. 
The results with the parameter choice \quad $c=10^2$, $\alpha = 10^{-6}$, $\beta = 6 \times 10^{-5}$ are as follows:  
The minimum of the potential is given by
$
S_{{\rm min}} = 2.23 \times 10^{-2}, \quad  
Y_{{\rm min}} = 1.12 \times 10^{-2}, \quad  
V(S_{{\rm min}},Y_{{\rm min}}) = 5.94 \times 10^{-12}.
$
The parameters of inflation are predicted as follows
\begin{eqnarray}
&&S_{{\rm end}} = 0.7394, \quad S_* = 10.90, \quad \mathcal{P_{R^*}} = 2.438 \times 10^{-9}, \nonumber \\
&&N = 58.79, \quad\,\,\, n_{S^*} = 0.9746, \quad \alpha_{S^*} = -4.303 \times 10^{-4}.
\end{eqnarray}
The Gravitino mass and the SUSY breaking scale are predicted as:
\begin{eqnarray}
m^{3/2} = | M_P e^{\frac{K}{2}} W | = 8.99 \times 10^{12} \,\, {\rm GeV}, \qquad 
F_S = 2.19 \times 10^{12} \,\, {\rm GeV}.
\end{eqnarray} 
We show the potential $V(S)$ minimized with respect to $Y$ in Fig. 1, and the evolution of the slow-roll parameters in Fig. 2. The stability of the potential minimum at $T=1$ can also be proved.\\
\begin{figure}[htbp]
\begin{minipage}{0.5\hsize}
\begin{center}
\includegraphics[scale=0.7]{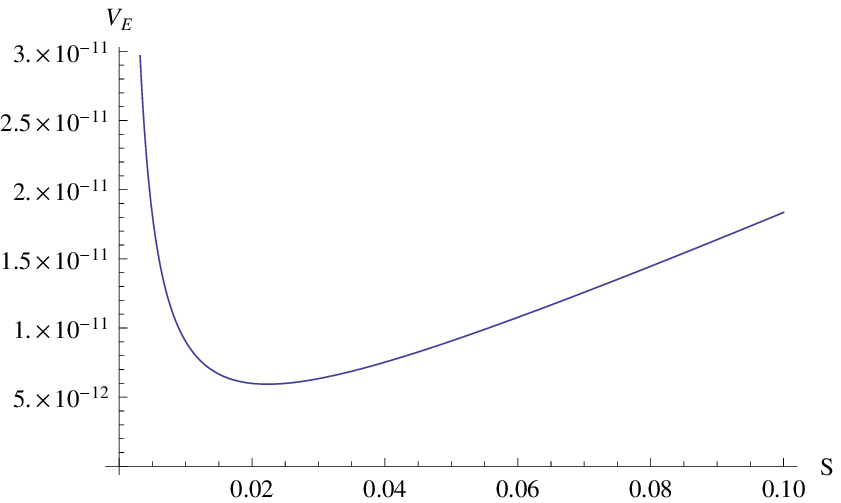}
\end{center}
\caption{\small The potential $V(S)$ minimized with respect to $Y$.}
\end{minipage}
\begin{minipage}{0.02\hsize}
\begin{center}
\end{center}
\end{minipage}
\begin{minipage}{0.5\hsize}
\begin{center}
\includegraphics[scale=0.55]{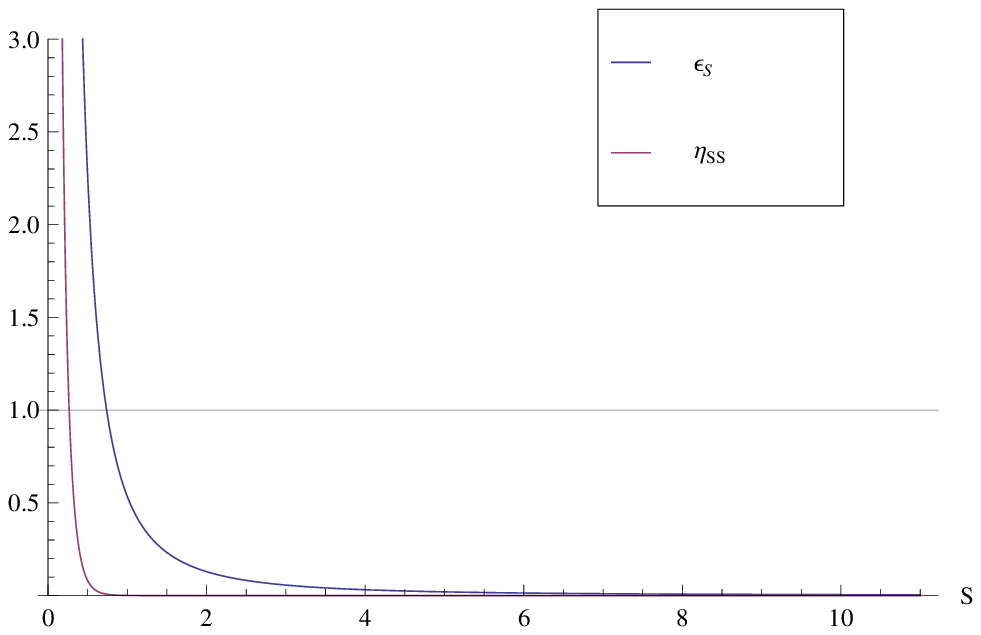}
\end{center}
\caption{\small The evolution of the slow-roll parameters. The blue curve represents $\epsilon_S$ while the red curve denotes $|\eta_{SS}|$. }
\end{minipage}
\end{figure}
This case seems to explain the WMAP observations well.
The slow-roll condition is well satisfied, and the $\eta$-problem can be resolved. 
Using the slow-roll approximation, the number of $e$-folds at which a co-moving scale $k$ crosses the Hubble scale $aH$ during inflation is given by:
$
N\sim -\int^{S_*}_{S_{\rm end}}\frac{V}{\partial V}dS \sim 58.72,
$
by integrating from $S_{\rm end}$ to $S_*$, fixing the parameters $c\ {\rm and}\ b$ as well as $\alpha\ {\rm and}\ \beta$. That is, our potential has the ability to produce the cosmologically plausible number of $e$-folds.\\
A scalar spectral index for a scale dependence of the spectrum of density perturbation and its tilt are defined by
$ n_s-1=\frac{d\ln \mathcal{P_R}}{d\ln k} $ and
$ \alpha_s=\frac{dn_s}{d\ln k} $.
Substituting $S_*$ into these formula, 
we have $n_{s*}\sim 0.9746$ and $\alpha_{s*}\sim -4.3 \times 10^{-4}$. 
Estimating the spectrum of the density perturbation $\mathcal{P_R}$ caused by slow-rolling dilaton, we found 
$\mathcal{P_R}_* \sim 2.438 \times 10^{-9}$.
Finally, the ratio $r$ between the scalar power spectrum $\mathcal{P_R}_*$ and the tensor one $\mathcal{P}_T$ is predicted as $r\sim 6.755 \times 10^{-2}$, which seems to be in the range possibly observed by the Planck satellite soon. 
The energy scale of the potential at the minimum, moreover, is given as 
$V\sim 5.9 \times 10^{-12}$, which is non-negative and may be considered to be small. 
It is the end of inflation, when one of the slow-roll parameters $\epsilon_\alpha$ or $\eta_{\alpha\beta}$ reaches the value 1. After passing through the minimum of the potential, reheating will begin.\\ 
Let us consider the Super Higgs mechanism in our model.
The inflatino field $\tilde{S}$ with its mass $m_{\tilde{S}}=0\ {\rm{GeV}}$, which is the SUSY partner of the inflaton (dilaton) field $S$, can play the role of the Higgsino field. Because the metric elements satisfy $g_{ST}=g_{SY}=0$ in the K\"{a}hler metric $g_{ij}$, $S$ does not mix with $Y,\ T$. Then the result of Super Higgs mechanism is given as 
\begin{eqnarray}
{\mathcal{L}}_{\rm{SHM1}}
=ee^{\frac{G}{2}}\Big\{\Big(\psi_\mu+\frac{i}{3\sqrt{2}}G_{S^*}\bar{\tilde{S}}\bar{\sigma}_\mu\Big)\sigma^{\mu\nu}\Big(\psi_\nu-\frac{i}{3\sqrt{2}}G_{S^*}\sigma_\nu \bar{\tilde{S}}\Big)+\frac
{1}{2}(G_{S^*S^*}+\frac{1}{3}G_{S^*}G_{S^*})\bar{\tilde{S}}\bar{\tilde{S}}\Big\}\label{shm}.
\end{eqnarray}
The last term of Eq.(\ref{shm}) implies the mass of $\tilde{S}$, which is proved to be exactly zero in our model. 
The first term can be identified with the mass term of the massive gravitino field, whose mass is given by $m_{3/2}=e^{G/2}$. 
This is the scenario of Super Higgs mechanism in our model. 
The predicted value of gravitino mass is given as:
$m^{3/2} = | M_P e^{\frac{K}{2}} W | = 8.99 \times 10^{12}$ GeV.
The scale of SUSY breaking is 
$F_S = 2.19 \times 10^{12} \,\, {\rm GeV}$.\\
After the scalars $S,Y,T$ are canonically normalized and the masses diagonalized, these masses are calculated as \\
$M_{S''}=9.98 \times 10^{12}$ GeV, $M_{Y"}=2.61 \times 10^{16}$ GeV, $M_{T''}=2.27 \times 10^{12}$ GeV, \\
where the mass eigenstates are denoted by $S'',Y'',T''$.\\
The decay rate of the process $Y'' \rightarrow \psi_{3/2}+\psi_{3/2}$ is estimated as
\begin{eqnarray}
\Gamma(Y'' \rightarrow \psi_{3/2}+\psi_{3/2}) = 4.78 \times 10^4 \,\, {\rm GeV}, \quad \quad
\tau (Y'' \rightarrow \psi_{3/2}+\psi_{3/2})  = 1.38 \times 10^{-29} \,\, {\rm sec}.
\end{eqnarray}
This process occurs almost instantly.\\
In order to estimate the reheating temperature, the decay rate of $S''$ into gauginos is calculated. 
By using the term 
${\mathcal{L}}_{gaugino}=\kappa \int d^2\theta f_{ab}(\phi)W_{\alpha}W^{\alpha}$, $f_{ab}(\phi)=\phi\delta_{ab}$, 
the interaction between $S(=\phi)$ and gauginos $\lambda^a$'s is given by
\begin{eqnarray}
&&{\mathcal{L}}_{gaugino}
=\frac{i}{2}f^R_{ab}(\phi)\left[\lambda^a\sigma^\mu\tilde{\mathcal{D}}_\mu\bar{\lambda}^b+\bar{\lambda}^a\sigma^\mu\tilde{\mathcal{D}}_\mu\lambda^b \right]-\frac{1}{2}f^I_{ab}(\phi)\tilde{\mathcal{D}}_\mu\left[\lambda^a\sigma^{\mu}\bar{\lambda}^b\right] \nonumber \\
&&\qquad -\frac{1}{4}\frac{\partial f_{ab}(\phi)}{\partial \phi}e^{K/2}G_{\phi\phi^*}D_{\phi^*}W^*\lambda^a\lambda^b +\frac{1}{4}\left(\frac{\partial f_{ab}(\phi)}{\partial \phi}\right)^*e^{K/2}G_{\phi\phi^*}D_{\phi}W\bar{\lambda}^a\bar{\lambda}^b. \label{gaugino_decay}
\end{eqnarray}
By using the relation $F_S \sim M_Pm_{SP}$, which holds for the mass of SUSY particles,
the gaugino masses can be estimated as 
\begin{equation}
m_\lambda = \frac{F_S^2}{M_P} \sim 1.97 \times 10^6 \,\, {\rm GeV}. 
\end{equation}
Then the decay rate of  $S'' \rightarrow \lambda+\lambda$ is given by
\begin{eqnarray}
\Gamma(S''\to \lambda\lambda) = 2.96 \times 10^{-3}  \,\, {\rm GeV}.
\end{eqnarray}
The reheating temperature $T_R({\rm gaugino})$ is derived from the Boltzmann equation by using the decay rate, and is given by
$
T_R({\rm gaugino})=\left(\frac{10}{g_*}\right)^\frac{1}{4}\sqrt{M_P~\Gamma(S'' \to \lambda + \lambda) },
$
where $g_*$ is the number of the effective degrees of freedom of MSSM, i.e.  $g_* =228.75$.
\\
By inserting the decay rate,  the reheating temperature is estimated as
\begin{eqnarray}
T_R({\rm gaugino}) = 3.88 \times 10^{7} \,\, {\rm GeV}. 
\end{eqnarray}
Because the reheating temperature is lower than the gravitino mass scale, gravitino reproduction will not occur after reheating.  
Though only one example of parameter choices has been  discussed here, 
the other seven candidates of parameter choices, which are compatible with WMAP data, have already been found by the authors. Some of the examples show that the gauginos can be observed by LHC experiments\cite{05}.

\end{document}